# Grand Technologies for Grand Energy Challenges
## A Futuristic Scenario for Solar Energy in the Age of Information

**Vahid Moosavi (svm@arch.ethz.ch)**

**Abstract**

Instead of getting involved in either extremes of dispute around climate change as one of our grand challenges, we opened the space of potential policy responses from a systemic view and showed why current climate change mitigation policies are not successful as planned. Further, as a potential futuristic scenario, neither a projection nor a prediction, we showed how solar based energy systems are different than other current energy systems and how we can conceive of them as grand technologies which dissolve the whole landscape of energy management in the age of information.

## Introduction

When, humankind learned how to release the stored sun-energy in woods into the fire and light, it was a great achievement for human society, but no one could imagine about deforestation at that time. Or when, people started using whale oil for lighting, no one could think of overfishing and the scarcity of the whales, but the economic force was more powerful than natural ecology of whales. The outcome was unimaginable. "…*Whaling was the fifth-largest industry in the United States; in 1853 alone, 8,000 whales were slaughtered for whale oil shipped to light lamps around the world*[1]" But thanks to human intellect, steam engine coupled by coal and oil, changed the whole story and it opened up a new era of industrialization, urbanization and mobilization. However, when genius Edison illuminated the New York City with a soft and continuous light, produced without gas or flame just by transmitting electricity and light bulbs, no one could imagine what unintended consequences of this new technology are as we see these days as ever-growing demand of electricity, generated mainly by fossil fuels (Figure 1).

---

[1] http://www.nytimes.com/2008/08/03/nyregion/03towns.html?_r=0

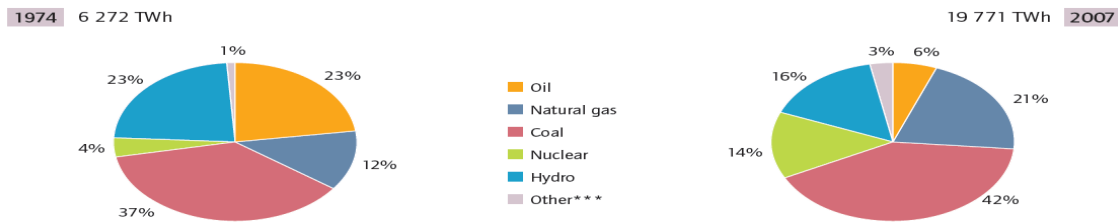

**Figure 1- Electricity generation by different fuels (IEA 2009)**

So, the point is that during the history, always successful technologies come with a grand proposal to dissolve the grand challenges, but later it turns out that they had some side effects, which is unknown in advance. However, it seems so far human society has been improving its conditions and every time a new technology arrives, it is better than its predecessor. Just as a big picture figure 2 shows how different indicators such as life expectancy, GDP per capita and population size have been increasing during the age of industrialization.

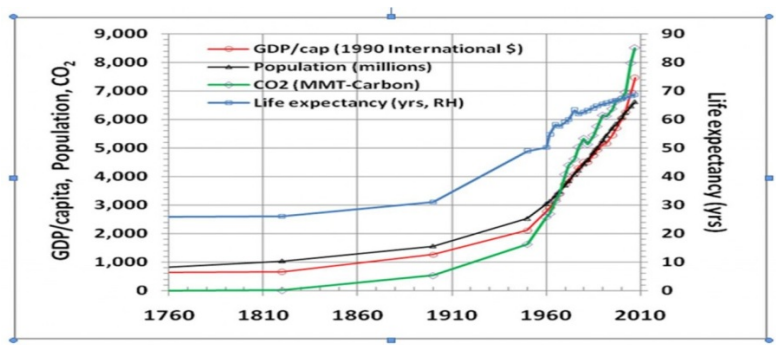

**Figure 2- World population (P), GDP per capita (A), life expectancy (LE), and CO2 emissions from fossil fuels in the Age of Industrialization[2].**

But nevertheless as we can see alongside these desirable trends, there are some undesirable consequences. As figure 2 shows there is a very strong correlation between $CO_2$ emissions with other desirable indicators. For sure, we should say these are all the results of industrialization, good and bad. Therefore, we are again facing a new challenge and the right question is how to deal with this problem in an appropriate way.

---

[2] Image from : http://www.masterresource.org/2010/04/population-consumption-carbon-emissions-and-human-well-being-in-the-age-of-industrialization-part-ii-a-reality-check-of-the-neo-malthusian-worldview/

Before discussing the main grand challenge of our days, *climate change and global warming*, it is worth to remind ourselves about the issue of "*natural resource depletion*" which took us for 25 years to realize that actually those kinds of pessimistic concerns about 2YK did not happen as for example it was predicted by "*limits to growth*" (Meadows, et al. 1972). Thanks to advancements in recent technologies such as "*shale gas*" and "*fracturing*", or possibilities in biofuels (Huber & Dale, 2009), it turned out that the issue of limited resources is not a purely geological limit, but more depending on the demand and the price of energy (McCabe 1998) as well as other possible technologies for energy production. In other words, it was a technological problem and in fact, it is our intellectual capabilities that define the limit of our technologies (Hovestadt et. al., forthcoming and Simon, 1998).

However, this time the challenge of global warming and climate change seems to be a serious issue. In fact, despite remarkable achievements, industrial age has had a serious unintended consequence, which is mainly due to over use of fossil-based energy systems. During the process of energy conversion (fuel combustion), fossil fuels such as oil, coal and gas emit $CO_2$ and other *greenhouse gases* which in principle are necessary part of our environment, but after certain level of atmospheric concentration of these gases, the whole eco-system will change and it causes an increase of the temperature in earth, which is called *global warming* (IPCC 2013). As it is shown in figure 3 during last 50 years, rapid growth of demand for fossil fuels has created a risky situation that might have irreversible harmful effects on environment.

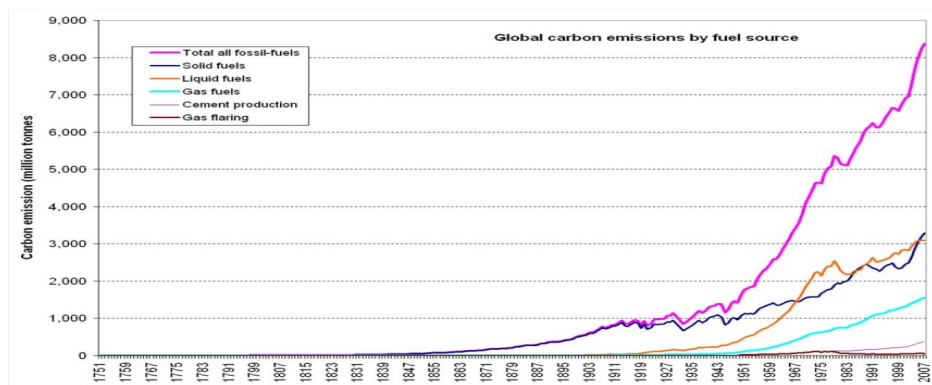

**Figure 3- Carbon emission from fossil fuels (Boden 2010)**

Although for a time there were lots of disputes and disagreements about the validity of the predictions about global warming, according to (Hasselmann & Barker, 2008): "*Today, the reality and widespread implications of anthropogenic climate change are no longer seriously disputed. Climate change has moved to the top of the political agenda. The question is no longer "Are we experiencing a human-made change in climate?" but "What must we do about it?"*". In this work, our focus is on the second question but in a positive way. According to (Hovestadt et. al., forthcoming)," *There are two possible stances to take. One is to say, here is a grave situation, we must deal with it as a matter of urgency and in order to deal with it we refer to and draw on everything we know. The other one is to say, here is an interesting situation. It opens up all manner of possibilities and we can deal with it in ways we've never even imagined before.*" We took the second approach.

## 2- How to Deal with the Problem of Climate Change or How to See New Possibilities (A Systemic View)

Energy systems are among complex systems which demand for a multi-perspective analysis of the situation. Therefore, it is not easy to identify one problem only around one aspect. For example, the only factor that seems to be important in most of the discussions around climate change is the negative effect of fossil fuels on our environment, while these, now a days dirty!, natural resources still have lots of great positive effects on our daily life (Figure 2). Therefore, when assessing different approaches it is important to have a holistic perspective to the situation, considering different aspects of the current energy landscape, including

- Technological and technical aspects
- Economic and financial aspects
- Political parties and geopolitical aspects
- Regulatory and governmental aspects (e.g. national energy security)
- Environmental aspects
- Health
- Social aspects and life style
- Transportation and urban aspects

From a systemic point of view according to (Ackoff, 1978), there are four main general approaches in dealing with a problem, namely: *Absolution, Resolution, Solution, and Dissolution.*

*Absolution* means when we see a problematic situation but we do not consider it and we hope it will be fine by itself or we think it is destined to happen and we are helpless about this phenomenon. *This is the state of denial or inability*. For example, for the problem of whale oil dependency and whale scarcity, the trend that was happening in the industry was mainly based on *absolution*, in which whale were close to extinction, but industries were hunting based on their growing demand.

*Resolution* is when we propose an action to a problem which is not necessarily the optimized solution but based on common sense or previous experience it might improve the problematic situation. *This is the consultants approach*.

Finding a *Solution* for a problem is coming originally from mathematics, in which the idea is that first we need a sufficient and complete (rational) model of the environment including the constraints imposed by the problematic situation to us and then based on a set of goals (objectives) one can find the best answers which optimize all of the defined goals and satisfies the constraints. *This approach is the act of rational modeling, optimization and planning.* For example, for the problem of whale oil, possible solutions could be to optimize the consumptions by finding advanced way of whale oil processing, or to find the best time for whale hunting and forcing the industry to follow a certain regulations, which demands for better understanding of ecological dynamics as well as coordination and obligatory rules for industries, governments and consumers.

Finally, *dissolution* is when instead of modeling the system and trying to satisfy the constraints, we change the whole problematic environments and dis-solve the problem instead of solving it. *This is normally the abstraction from the problem space and creation a new space.* Most of technological innovations are among dissolutions. In the case of whale oil or deforestation due to overconsumptions of wood, the dissolution was the invention of steam engine with introduction of fossil fuels which changed the whole discussion space to another landscape. Normally, after *dissolution* people forget the previous problem and the next generation could not imagine that one day we had this problem, but in the case of *solution* always we need to establish monitoring and control.

Based on these four categories of treating problems, it is possible to analyze the continuum of current approaches to the problem of climate change. Table 1, shows different policy responses to the problem of climate change based on the four above mentioned scenarios.

Table 1- Different approaches in dealing wiht climte change problem

| Approach | Climate change | Strategy |
|---|---|---|
| Absolution | Denial of climate change | Inactive or Passive |
| Resolution | Geo-engineering and Negative emission (IEA 2013) | Reactive |
| Solution | **(Optimize the behaviors in the current game to minimize the loss)** Climate change mitigation policies: (Hasselmann & Barker, 2008): <br> • Demand reduction <br> • Energy efficiency <br> • Carbon tax <br> • Carbon trading (Hepburn, 2007) | Reactive |
| Dissolution | **(Propose a new game that creates a new landscape)** Pushing technological innovation toward clean and abundant energies such as PV and changing the paradigm of infrastructural thinking and centralized planning (section 5) | Proactive |

So based on what mentioned above, in *absolution* based approaches, either we deny the problem or we think it is not possible to change the climate change and global warming, but instead we take a passive strategy toward the coming problem. For the case of *resolution* and *Solution* in general the strategy is reactive, in which we perceive an undesirable situation (i.e. $CO_2$ emissions) and we directly react to these problem either by neutralizing the effect (IEA 2013) or by opposing it (i.e. demand reduction strategies) or by playing with the problematic situation (i.e. energy efficiency and carbon trading (Hepburn, 2007), but as we explained before, still we are thinking in the same paradigm and space that created the problem. The first main argument in this work is that *solution* approaches conceptually cannot solve the problem of climate change as we will discuss in the next section.

The final approach, *Dissolution,* aims to be proactive, and to change the whole game instead of winning the game. In this work we try to show how solar based energy systems can potentially change the future of our energy toward a clean environments based on an abundant energy source from sun.

*Note:* As an important note we should mention that the differences between two main approaches of *climate change mitigation* and *trusting on human intellect and technology development* is an old dispute between the so called "Malthusians" after the work of Thomas Malthus on one side and "Cornocopians" on the other side, in different issues such as sustainability, global warming, overpopulation, peak oil, ozone depletion, deforestation and so on. As famous examples of Malthusians one can refer to "*The Population Bomb*" (Ehrlich, 1971), which was not a correct forecast or the predictions by "*the limits to growth*" (Meadows, et al. 1972), that did not happen by the end of 20$^{th}$ century. From famous cornucopian works, one can refer to *The Ultimate Resource* (Simon, 1998), which argues that growth is, unlimited and that human ingenuity is "the ultimate resource."[3]

Although in this work our position is toward cornucopian approaches, we think instead of too much focus on the limits of the other approach, it is important to have a constructive and positive investigation on the possibility of having a prosperous future of energy.

*Note:* Beside, although as it can be seen in most of political disputes, two main approaches are in two extreme sides of the solution continuum, but in this work we think instead of having a dichotomist view (either-this-or-that) it is possible to take a dialectic view to all the solutions. And in fact, the beauty of the current dynamic in political discussions is its diversity as a whole, but as an individual opinion, we need to take a position.

## 3- Climate Change Mitigation Policies or How to Play the Current Game Better?

"***Problems cannot be solved at the same level of awareness that created them***." - Albert Einstein

Based on (IEA 2013) "*Despite the insufficiency of global action to date, limiting the global temperature rise to 2 °C remains still technically feasible, though it is extremely challenging.*" Or "*Carbon pricing is gradually becoming established, and yet … consumption of coal, the most carbon-intensive fossil fuel, continues to increase globally.*"

---

[3] http://rationalwiki.org/wiki/Cornucopian_vs._Malthusian_debate

"*Action to improve energy efficiency is increasing, but two-thirds of the potential remains untapped*" (IEA 2012). Although the language of majority of reports on climate change-mitigation policies are not negative and sometimes they might report the decrease of $CO_2$ emissions (IEA 2013), but figure 4 shows how GDP growth is correlated by $CO_2$ emission. This might show how effective climate change mitigation policies are and what would be the effect of directly pushing $CO_2$ down.

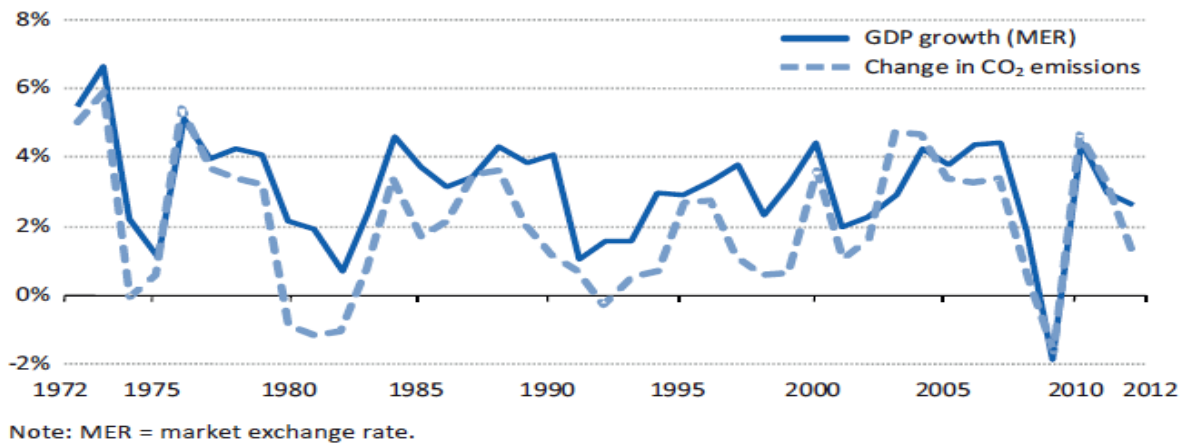

Figure 4- Growth in global GDP and in energy-related CO2 emissions (IEA 2013)

However, there are reports believe the underlying reasons for unsatisfying results of these policies are not external, but internal factors within the community (Hasselmann & Barker, 2008). Factors such as lack of enough climate experts and their connections to political parties or more importantly lack of an "*international integrated assessing model*" are among the main reasons of failure. Therefore, more developed (i.e. complicated) models and simulations, which consider different countries, different sectors and a wider variety of different objectives is needed (Hasselmann & Barker, 2008). Here, our main argument is that this way cannot lead to a fruitful model and consequently an interesting outcome. Going back to main abstract categories of problem treatment, this is a *solution* seeking approach in a complex environment which theoretically does not succeed. As it is shown in figure 4 $CO_2$ emission is in coexistence with GDP growth showing several other development factors (Figure 2). Then, one must consider powerful forces opposite to majority of climate change mitigation solutions. And since our globalized economy is still based on the industrial paradigm, these kind of side effects (here $CO_2$ emission) is uncontrollable till we are living in this paradigm.

Following are some of other reasons of the failure of climate change mitigation policies:

- *Global-state vs. national-state*: Distribution of natural resources is not uniform all over the world, and only few countries have the privileges of earning money by selling fossil fuels, while the issue of climate change is a global problem. Therefore there is a contradiction between national benefits and global benefits.
- *Generic and bulk production of energy* historically has created a political structure of power, (i.e. by main international oil producers) which is out of the rich of any country and community to be influenced by. For example, the conclusion by the documentary switch[4] shows this paradox. On one side it mentions that meeting the future demand of energy based on growing demand is a challenging issue. Therefore, he invites people as individuals for energy saving, but at the same time he shows giant oil producer companies which are waiting for higher demand and therefore higher prices of oil makes it feasible for them to explore more. Here again we have two opposite forces, which forces from producers seem to be dominant.
- *Limits of model based planning in complex socio-technical systems*: In theory of modeling, the basic assumption behind the concept of planning and optimization is that we have a sufficient and complete structural and behavioral model of the real environment and then based on the validated model, selected goals and objectives, one can do the act of planning to find the optimum allocation of resources. However, theoretically mathematician Godel in his incompleteness theorem[5] shows the limits of analytical modeling in 1931. Further, due to nature of complex systems[6] such as energy landscape it is not possible to develop a comprehensive model and simulation of the real environment without simplifications. Although, this way of large scale modeling in socio-technical systems has been criticized widely in 1970s (Lee, 1973, and Ackoff, 1991), but still due to availability of computer power, there are some endeavors for such models for climate change mitigation policies(Weitzman,2008 and Weyant 2000) . Therefore, we see the similar complains (Hasselmann & Barker, 2008) about current climate models which were mentioned in 1970s (Lee, 1973):

- Hyper-comprehensiveness (too many purposes in one model)
- Hungriness (always data is not sufficient)
- Grossness
- Wrong-headedness (always wrong predictions)
- Complicatedness
- Mechanicalness
- Expensiveness

---

[4] http://www.switchenergyproject.com/
[5] http://en.wikipedia.org/wiki/G%C3%B6del's_incompleteness_theorems
[6] Complexity here is considered as a function of number of elements, interdependencies between elements and number of stakeholders in time

Therefore, we think the climate change mitigation policies as the current situation as *solution* approach to a complex problem cannot succeed, as the produced models and policies are all in the same level of industrial and rational thinking which has created the problem itself. Based on the quote from Einstein in the beginning of this section, we think a proactive dissolution is required which goes beyond the current concept of planning, governance and economy.   We think now at the age of information, world needs to approach the problem of climate change in a different way, and instead of trying to find the optimum solution in the current game, we need to introduce a new game.

## 4-Solar Technology or How to Introduce a New Game?

Currently in majority of  reports and discussions on energy sources, solar is a small part of renewables and since the whole contribution of renewables is very small comparing to traditional fossil fuels and others like nuclear, in most of current energy discussions, solar is almost absent. First of all, we want to reflect on the division of energy sources as renewable and non-renewable, which is mainly based on old paradigm of natural resource depletion, while nowadays the question of resource limits is not of high importance, as now it is possible to produce petrol out of food and biomass (Huber & Dale, 2009) which is a renewable source of energy, while classically, oil is among non-renewables. Therefore, the issue of renewability is relative. At the same time, considering the challenges of fossil fuels, and safety issues of nuclear power, there are growing numbers of works, dedicated on future energy with focus on solar based energy systems (IEA, 2011). The current view on PV panels and other solar energy systems is that 1-they are in principle abundant, but 2- their efficiency rate for capturing the sun energy is very low and 3- due to several other reasons (Bony, et. al. 2010) their produced energy is still expensive comparing to other sources of energy. 4-Another limit of PV electricity is the problem of energy storage, but of course grid-connected PV systems or in the case of off-grid PV, new advanced storage systems are the solutions.

However, as we discussed in previous section, regardless of the current technical problems we think solar based energy systems have specific potentials that in long term not only can promote the issues of $CO_2$ emission and energy supply in an economic way, but they can open up new ways of thinking about energy and energy planning.

From a multi-perspective view to energy systems, the following differences can be identified.

- *Abundant source of clean energy*: If humanity could capture with 10% efficiency, only one percent of the solar energy striking the earth we would have access to six times as much energy as we consume in all forms today, with almost no greenhouse gas emissions.(Naam, 2011)
- PV is about *capturing* the potential not the *consumption of the source of energy*.
- *The efficiency rate of PV* panels is up to the technology growth which seems exponential, while traditional energies are limited to laws of physics for energy conversion, like thermodynamics laws in fossil fuel combustion.
- *Geographic Distribution of solar energy* is fairly uniform all over the world, unlike most of traditional energy sources.
- *Distributed electricity generation* takes place instead of *bulk generation*.
- *Application Oriented electricity generation*: electricity generation by PV can be application oriented while classically, energy is being produced in a generic way. Therefore, according (Bazilian et. al. 2013) economic measures such as grid parity might not be appropriate for

- comparing PVs with other bulk based energy productions, since grid parity assumes the wholesale energy provision, while PV can be implemented in very small scale.
- *Ownership* of energy and energy production goes from national or giant private companies to an individual level, which will change the whole paradigm of energy planning and energy tariffs.
- *Going beyond the concept of rational (long term) planning and centralized decision making.* Similar to IT industry which only supports the provision of equipment in a competitive market, there is no need to classic national and regional energy planning systems.
- *Shift from infrastructural thinking*: Comparing to traditional energy networks which always need heavy infrastructures, PV can be implemented with light infrastructural requirements.

However, one can say that despite all of these different aspects of solar technology, still it is not there!! Yes, it is correct, but soon or late it happens and one of the main goals of this study is to open up the discussion around different approaches in climate change policy and to provide a positive *dissolution* approach to the problem of climate change and energy management. This can attract the attention of policy makers and governments from a negative and crisis-based view to a positive and optimistic view of future.

In the next step, we will elaborate on the potentials of solar technology and its similarity to information technology. This analogy will draw a potential future of solar energy.

## 5- A Potential Futuristic Scenario (Energy as Information)

Regarding the future of technologies, in general there are three main approaches including *projection*, *prediction* or *forecasting* and finally *scenarios* or *stories*. Projections always think linearly and based on the current status of technology and demand they calculate the future capacity of the technology. Definitely, based on the current capacity of solar system, these projections do not show a considerable role for solar in future, but in a way they are just simple arithmetic, in which the underlying assumption is that the system is always the same as before. Second group is prediction and forecasting, which is more complicated than projections, but as we discussed before, due to complexity of the energy systems, it is theoretically impossible to predict something for future with high probability. The third approach is scenarios which are based on possible outcomes. This approach normally goes to the causes and relations between different elements and in tells a potential futuristic story, which is possible, but not necessarily destined to happen. This approach mainly calls for open discussions and the underlying belief is that the final future outcome is the emergent outcome of different individual actors. In this study we took the third approach and therefore, what is proposed is more based on the foundations and similar experiences that humankind has had historically.

As Paul Krugman says[7]: "*For decades the story of technology has been dominated, in the popular mind and to a large extent in reality, by computing and the things you can do with it …from faxes to Facebook, … [but] Our mastery of the material world, on the other hand, has advanced much more slowly. The sources of energy, the way we move stuff around, are much the same as they were a generation ago.*"

However, based on the above mentioned points about solar energy systems, we think that we are in a transition from industrial age to information age of energy.

According to (Hovestadt et. al., forthcoming) solar PV technologies are analogically similar to information technology and they are coming together based on three main reasons:

- *We get more energy by using more*: Traditionally energy is produced by *consuming* a resource, like burning the coal or wood, but information as a kind of resource has a self-reinforcing growth mechanism in which by using more information we are able to produce more information. Similarly considering the abundance of solar energy, the electricity generation in PV is not by consuming a resource but by using the material that has turned into PV cell. Therefore assuming the long life of these cells and their recycling process, we have more energy by using more, which is in contrast with the classic paradigm of natural resources.
- *Moore's law*: Information technology based products are following the famous Moore's law (Moore 1965), which shows how exponentially things get cheaper during the time. PV panels also as pure technological devices are seemingly following the Moore's law (Naam, 2011) (Figure 5). Therefore, regardless of how fast this exponential growth is happening, after a threshold PV based electricity gets much cheaper and then in that point regarding the abundant source of energy, a new dynamics will start, which is out of scope of energy efficiency, resource planning and so on, similar to what is now happening to flow of information. For example, how cell phone companies are finding new markets in those underdeveloped areas in Africa that are still waiting for basic infrastructural needs such water and electricity.

---
[7] http://www.nytimes.com/2011/11/07/opinion/krugman-here-comes-solar-energy.html?_r=3&hp&

- *Network*: Classic energy networks are hierarchical, one directional, and are based on one to many relations (one producers and many consumers) while information networks and potentially grid connected solar PVs are flat, multi-directional and based on many to many relations, in which similar to information networks (i.e. internet) a dynamic and complex environment will start to happen, which goes beyond the concept of rational and centralized planning in current energy networks.

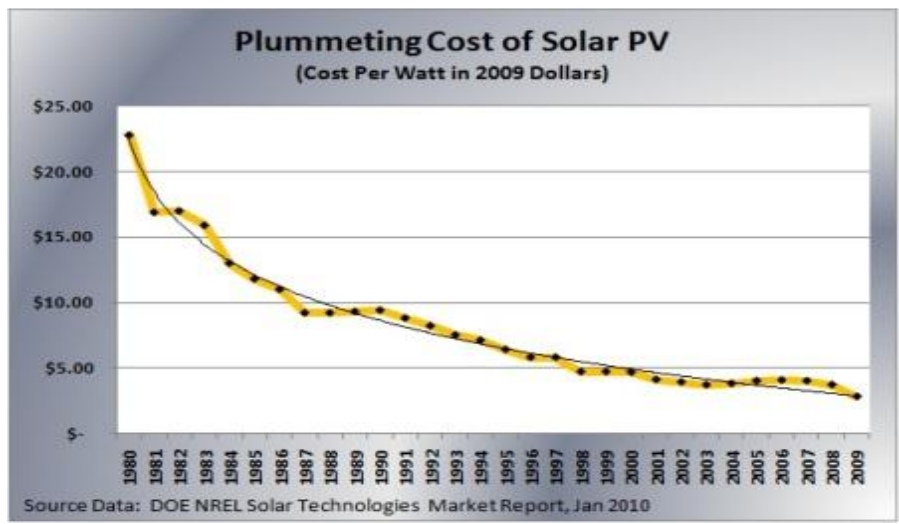

**Figure 5- Exponential rate of change in PV technology (Naam, 2011)**

Therefore, based on this futuristic scenario it is clear how the future of energy based on solar systems can be different than the current discussions on climate change mitigation solutions and their future scenarios. Further, it is expectable that with the paradigm of energy as information how the concept of energy will be decoupled from the land and physical space as it has already happened by information technology.

But nevertheless as the new technology will change our world to another world, we should eagerly expect the new challenges which are unknown yet.

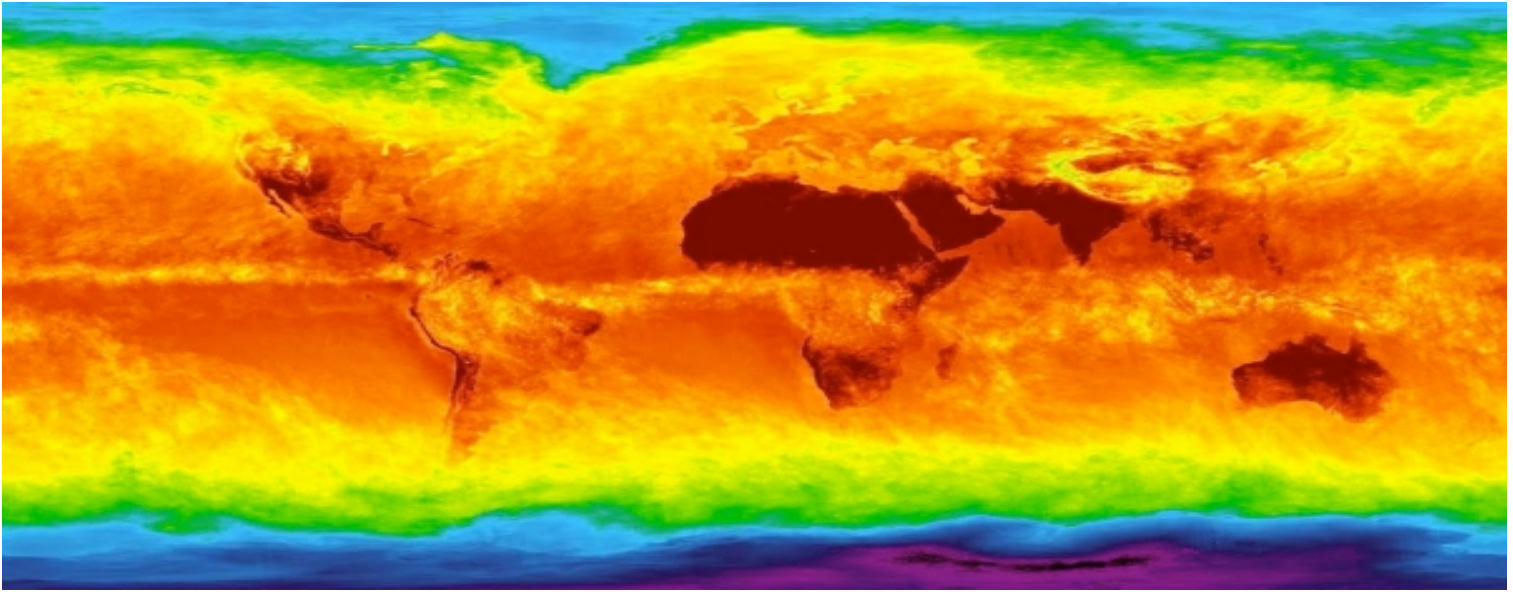

Figure 6- World average solar insolation, with red representing high solar insolation. Source: NASA

## References

1. (IPCC, 2013): Working Group I (WGI) contribution to the IPCC Fifth Assessment Report: approved Summary for Policymakers, p.2 (footnote 1) and p.12.
2. (IEA 2009): IEA Scoreboard 2009: 35 Key Energy Trends over 35 Years. International Energy Agency, 2009.
3. (IEA, 2011) Technology Roadmap: Solar photovoltaic energy, International Energy Agency, 2011.
4. (IEA 2012) *World Energy Outlook 2012*, OECD/IEA, Paris.
5. (IEA 2013) World Energy Outlook Special Report 2013: Redrawing the Energy Climate Map, International Energy Agency, 2013
6. (Meadows, et al. 1972)*The Limits to Growth: A Report to The Club of Rome (1972)*. Universe Books, New York, 1972.
7. (Huber & Dale, 2009) Huber, G. W., & Dale, B. E.. Grassoline at the pump. *Scientific American*, *301*(1), 52-59. 2009.
8. (McCabe, 1998). Energy resources-Cornucopia or empty barrel?. *AAPG Bulletin-American Association of Petroleum Geologists*, *82*(11), 2110-2134.

9. (Hovestadt et. al., forthcoming) Ludger Hovestadt and Vera Bühlmann with Sebastian Michael, The Genius Planet: Energy Scarcity to Energy Abundance (forthcoming).
10. (Simon,1998) Simon, J. L. The ultimate resource II: People, materials, and environment. 1998
11. (Boden 2010) Boden, T.A., G. Marland, and R.J. Andres. 2010. Global, Regional, and National Fossil-Fuel $CO_2$ Emissions. Carbon Dioxide Information Analysis Center, Oak Ridge National Laboratory, U.S. Department of Energy, Oak Ridge, Tenn., U.S.A. doi 10.3334/CDIAC/00001_V2010http://cdiac.ornl.gov/trends/emis/tre_glob.html
12. (Hasselmann & Barker, 2008) Hasselmann, K., & Barker, T. The Stern Review and the IPCC fourth assessment report: implications for interaction between policymakers and climate experts. An editorial essay. *Climatic Change*, *89*(3), 219-229. 2008.
13. (Ackoff, 1978 ) Ackoff, R. L., The Art of Problem Solving, John Wiley & Sons, New York, 1978.
14. (Ackoff, 1991) Ackoff, R. L., "The Future of Operational Research is Past," Critical SystemsThinking Directed Readings, R. L. Flood and M. C. Jackson (Editors),1991.
15. Hepburn, C. (2007). Carbon trading: a review of the Kyoto mechanisms. Annu. Rev. Environ. Resour., 32, 375-393.
16. Ehrlich, P. R. (1971). The population bomb (Vol. 68). New York: Ballantine Books. 1971
17. Lee Jr, D. B. (1973). Requiem for large-scale models. Journal of the American Institute of Planners, 39(3), 163-178.
18. Weitzman ML (2008) on modeling and interpreting the economics of catastrophic climate change.
19. Weyant JP (2000) An introduction to the economics of climate change. PEW centre on global climate change. Batelle, Washington DC
20. (Bony, et. al. 2010) Lionel Bony, Stephen Doig, Chris Hart, Eric Maurer, Sam Newman Achieving Low-Cost Solar PV: Industry Workshop Recommendations for Near-Term Balance of System Cost Reductions, Rocky Mountain Institute | RMI.org, 2010.
21. Naam, R. (2011). Smaller, cheaper, faster: Does Moore's law apply to solar cells?. Scientific American.
22. (Moore 1965) Moore, Gordon E. (1965). "Cramming more components onto integrated circuits", Electronics Magazine. p. 4. Retrieved 2006-11-11.
23. (Bazilian et. al. 2013) Bazilian, M., Onyeji, I., Liebreich, M., MacGill, I., Chase, J., Shah, J., ... & Zhengrong, S. (2013). Re-considering the economics of photovoltaic power. Renewable Energy, 53, 329-338.

Further points

http://foodtank.org/news/2013/04/five-ways-cell-phones-are-changing-agriculture-in-africa